\documentstyle[preprint,aps]{revtex}
\begin{document}
\draft
\title{Chaotic Interaction of Langmuir Solitons and Long Wavelength Radiation}
\author{R. Erichsen, G.I. de Oliveira, and F.B. Rizzato}
\address{Instituto de F\'{\i}sica - Universidade Federal do Rio Grande 
do Sul, Caixa Postal 15051, \\ 91501-970 \\ Porto Alegre, Rio Grande do 
Sul, Brazil}
\maketitle
\date{\today}
\begin{abstract}

In this work we analyzed the interaction of isolated solitary structures 
and ion-acoustic radiation. If the radiation amplitude is small solitary 
structures persists, but when the amplitude grows energy transfer towards 
small spatial scales occurs. We show that transfer is particularly fast 
when a fixed point of a low dimensional model is destroyed. 
  
\end{abstract}
%
\pacs{05.45.+b}
%
\section{Introduction}

\noindent Langmuir turbulence has been one of the most studied problems 
in modern nonlinear plasma physics. Over the last years a great deal of 
effort has been directed to its analysis, as well as to the analysis of 
related subjects as soliton dynamics, collapse, nucleation of cavitons, 
electromagnetic emission, and others \cite{dter78}. More 
recently, attempts have been made to understand the turbulence in terms 
of concepts of nonlinear dynamics and chaos 
\cite{dool85,moon90,xte95,gol95,olig96}.

The conservative version of Langmuir turbulence is described by the 
Zakharov equations that couple the slowly varying amplitude of a 
high-frequency electric field, the Langmuir field, to slow density 
fluctuations, the ion-acoustic field. Decay processes deposit energy 
into Langmuir fluctuations with long wavelengths and if the energy thus 
accumulated exceeds the threshold for modulational instability, solitons 
can be formed.

In addition to solitons a certain amount of ion-acoustic radiation is 
also generated, a fact that creates the possibility of nonlinear wave 
interaction involving these two types of structures: solitons of the 
Langmuir field and long wavelength ion-acoustic radiation. In more specific 
terms, what happens is that as solitons are formed their shapes 
exhibit temporal oscillations \cite{riz98}; if ion-acoustic fluctuations are 
also present, the possibility exists of interaction between the 
oscillatory degrees-of-freedom of solitons and the oscillating ion-acoustic 
waves. One has two length scales in the region of long scales. One of them 
is the soliton length scale, we shall call it $L_s$, and the other is the length 
scale of the ion-acoustic fluctuations, $L_i$. It has been shown that depending 
on the general conditions of the system, the mentioned interaction may lead to 
intense energy transfer from the spectral region of long wavelengths to the 
region with much shorter wavelengths, we call it $L_{sh}$ with 
$L_{sh} \ll L_s,L_i$. As energy moves into modes with small 
wavelengths, dissipation becomes progressively more important. However, as 
we are interested only in nonlinear transfer processes, we discard 
dissipation in a first approximation. It has been suggested that energy 
transfer occurs when the interaction is of chaotic nature. Presumably the 
process underlying the transfer is related to the diffusive processes induced 
by the presence of a stochastic drive in the system; the stochastic drive 
would be formed by the chaotic degrees-of-freedom \cite{lili91}.

Now we come to our point. In various earlier simulations 
\cite{olig96,riz98,meimei87} a modulationally 
perturbed plane wave is launched into the system. If the system is unstable 
a number of solitons and additional ion-acoustic radiation are formed. 
Solitons interact with each other and with the radiation, and transfer 
of energy towards small spatial scales $L_{sh}$ may take place if 
nonintegrable features are prominent. The problem here is that this kind of 
simulations does not examine properly the interaction of individual solitons 
and the radiation, since soliton-soliton collisional processes cannot be 
disregarded under such conditions. It is not even clear which type of 
interaction, if soliton-soliton or soliton-radiation, is the dominant 
responsible for the transfer. In addition, several systems display a small 
soliton density that collisions are unlikely - in these systems one should 
focus attention on the individual interaction involving one single 
oscillating soliton and ion-acoustic waves. This is the purpose of the 
present paper. We shall examine the system evolving from an initial condition 
where only one single oscillating soliton and some radiation are present. In 
addition to simulations we develop a model where we perform averages over 
fast variables in order to make estimates with regard to the behavior of the 
collective variables of the system. 

As will become clear, energy transfer starts to take place when the 
collective variables become chaotic. 
In general we shall see that while for moderately and small amplitudes of the 
perturbing ion-acoustic radiation solitons can 
be at least seen as metastable structures in the system, for large 
amplitudes transfer is fast and initial solitary structures are 
rapidly destroyed. It has been argued that soliton are robust enough to describe 
final states of this type of system \cite{deg74,kry80,dya89,isi94}. But what we 
see here is that even if some solitons are present in 
asymptotic states of large amplitude regimes, those solitons are not the same 
present in earlier times - after the initial solitons are destroyed there are 
long stretches of times over which no organized structures are seen. 

We finally mention that a number of works have already analyzed the interaction 
of localized structures and perturbations with longer wavelengths. In some of 
them only low dimensional models were investigated \cite{mal97}, and in 
others where full simulations were performed, chaotic dynamics was not the 
issue, although some nonintegrable features like soliton fusion have been 
reported \cite{meimei87}.

We organize the paper as follows: in \S 2 we introduce the basic model and 
the numerical techniques to be used here; in \S 3 we discuss our initial 
conditions and perform the appropriate averages to single out the relevant 
collective variables; in \S 4 we compare the low dimensional model with full 
simulations, and in \S 5 we summarize the work.  

\section{Basic equations and numerical techniques}
 
\noindent The one dimensional Zakharov equations governing the Langmuir 
turbulence can be written in the adimensional form \cite{olig96}
\begin{equation}
i \partial_t E + \partial_x^2 E = n \ E, 
\label{cael}
\end{equation}
\begin{equation}
\partial_t^2 n - \partial_x^2 n = \partial_x^2 |E|^2,
\label{caion}
\end{equation}
with $\partial_t \equiv \partial/\partial t$, 
$\partial_x \equiv \partial / \partial x$. $E(x,t)$ is the slowly 
varying amplitude of the high-frequency Langmuir field and $n(x,t)$ are slow 
density fluctuations associated with the ion-acoustic field. The NLS equation 
\begin{equation}
i \partial_t E + \partial_x^2 E + |E|^2 \ E = 0 
\label{nls}
\end{equation}
is obtained from the set (\ref{cael}), (\ref{caion}) if one is allowed to 
approximate Eq. (\ref{caion}) in order to replace $n$ with $-|E|^2 + const.$ 
This approximation is called subsonic because it requires very slow time 
scales that $\partial_t^2 \, n(x,t) \ll \partial_x^2 \, n(x,t)$.

Our numerical approach is based on a pseudo-spectral method. We assume spatial 
periodicity with basic length $L$ and expand $E(x,t)$ and $n(x,t)$ into 
Fourier series as
$$
E(x,t) = \sum_{m=-{N \over 2}}^{+{N \over 2}} E_m(t) e^{i\,mkx} 
\rm{, \> \> \> and}$$
\begin{equation} 
n(x,t) = \sum_{m=-{N \over 2}}^{+{N \over 2}} n_m(t) e^{i\,mkx}.
\label{eexp}
\end{equation}
The basic wavevector is defined in terms of the system length $L$ as 
$k=2 \pi / L$, and the integer $N$ represents the 
number of modes used in the simulations. To represent a continuous 
system one should take the limit $N \rightarrow \infty$. In practice we let 
$N = 1024$, removing half of the modes to cure aliasing 
problems associated with the FFT routines. Comparisons with $N = 2048$ 
indicates numerical convergence in terms of number of modes. Accuracy 
is further checked by varying the tolerance factor of the numerical 
integrator and by monitoring the conserved energy \cite{dter78}. We find that 
relative fluctuations in energy are about one part in $10^6-10^8$ and that 
variations of tolerance factor do not produce alterations in the outcome of 
runs. 

Solitons of amplitude $\sqrt{2}\,a_s$ are formed when an homogeneous train of 
Langmuir radiation of amplitude $|E_o|^2 \sim a_s$ becomes modulationally 
unstable. The subsonic growth rate $\Gamma$ for a perturbation with 
wavevector $k$ superimposed on the homogeneous train can be estimated as
\begin{equation}
{\Gamma \over k} \sim \sqrt{|E_o|^2 - k^2}.
\label{growth}
\end{equation}
From relation (\ref{growth}) one sees that the only unstable modes are those 
for which $|E_o|^2 > k^2$. Now when $|E_o|^2 - k^2 \ll 1$, $\Gamma \ll k$. If 
this condition holds for the majority of modes, $\partial_t \ll \partial_x$, 
ion-acoustic fluctuations are mostly enslaved to the Langmuir field, and 
approximation (\ref{nls}) can be used. On the other hand when $|E_o|^2$ is not 
exceedingly small ion-acoustic fluctuations with $k \ll |E_o|$ may not be 
completely enslaved to the Langmuir field. Those free fluctuations are to be 
seen as independent degrees-of-freedom whose presence is capable of destroying 
the integrability of the system. Given that the maximum growth rate occurs for 
$k_{max} \sim |E_o|$ and that the typical length scale of a soliton is 
given by $L_s \sim 1/k_{max}$, free ion-acoustic radiation of wavevector 
$k_i = 1/L_i$ typically appears in the spectral region for which 
\begin{equation}
L_i \gg L_s.
\label{relcomp}
\end{equation}

\section{Collective variables and low dimensional model}

\noindent Our system is multidimensional but we would like to see whether 
a small subgroup of modes is more active than the remaining. If this 
is the case one could try to describe the basic features of the full 
dynamics by a low dimensional approximation. As it turns out, such an 
approximation appears to be possible.  

To see how to obtain the low dimensional model, we proceed as follows. We 
first recall that as initial conditions we are interested in 
configurations with isolated solitary structures. To represent this sort of 
states either analytically or in the simulations we shall first 
determine the stationary one-soliton-solution for the full problem. We 
start by taking $\partial_t= 0$ in Eqs. (\ref{cael}) and (\ref{caion}) 
from which we first get $n \approx - |E_s|^2 + {\rm constant}.$ Substituting 
this relation into Eq. (\ref{cael}), after some algebra one obtains
\begin{equation}
E_s(x) = \sqrt{2} \> \xi \> sech (\xi \, x),
\label{shapest}
\end{equation}
which is the expression we are looking for. $\xi$ is an arbitrary factor 
that measures either the amplitude or the inverse width of the soliton. We 
point out that due to the nonlinearities and dispersion of the problem 
a precise balance between amplitude and width is needed. If we call 
$a_s \equiv \xi$ and $w_s \equiv {1 \over \xi}$, it is indeed seen that the 
following relation holds: 
\begin{equation}
a_s = {1 \over w_s}.
\label{width}
\end{equation}

We had mentioned that our interest is to see what could happen with 
the soliton when it starts to interact with free ion-acoustic radiation. 
Based on several results one knows already that the basic soliton 
solution must be allowed to display temporal 
oscillations. The problem now is how to describe those oscillations 
in a compact way. And the answer is known: one first write down an ansatz 
solution for the soliton field where amplitude is however not 
correlated to the width according to the static relation (\ref{width}). 
The ansatz solution is therefore generically written in the form
\begin{equation}
E(x,t) = \sqrt{2} \> a(t) \> sech \, ( {x \over w(t)} ) \> e^{i \Phi(t)},
\label{shape}
\end{equation}
where $a(t)$, $w(t)$, and $\Phi(t)$ are all unknown as yet. The phase 
factor $\Phi$ is included to incorporate the complex structure of the 
solutions of the set (\ref{cael}),(\ref{caion}). As for the ion-acoustic 
field interacting with the soliton field, one writes 
\begin{equation}
n(x,t) = - |E(x,t)|^2 + (A(t) \> e^{ikx} + {\rm c.c.})
\label{shapeion}
\end{equation}
Here we write the ion-acoustic field as a sum of the pure adiabatic response to 
the soliton field, plus some free radiation that will actually interact with 
the isolated nonlinear structure. $A(t)$ is the amplitude of the radiation 
field and c.c. stands for complex conjugate. The next step is to derive 
the appropriate governing equations for the four time dependent parameters, 
$a(t),w(t),\Phi(t)$, and $A(t)$. This is more easily done with help of average 
Lagrangean techniques. The full Lagrangean from which one obtains the original 
set (\ref{cael}),(\ref{caion}) reads  
\begin{equation}
L = \int {\cal L} dx \equiv \int [ 
{i \over 2} (E^\ast \partial_t E - E \partial_t E^\ast) - 
|\partial_x E|^2 - |E|^2 \partial_x \nu + {1 \over 2} [(\partial_t \nu)^2 - 
(\partial_x \nu)^2] ] dx, 
\label{lagf}
\end{equation}
where the dynamical variable $\nu(x,t)$ is introduced in the form 
$n(x,t) \equiv \partial_x \nu(x,t)$. The Euler-Lagrange equation for 
$E(x,t)$, for instance, is written as 
\begin{equation}
\partial_t {\partial {\cal L} \over \partial (\partial_t E)} = 
{\partial {\cal L} \over \partial E} - 
\partial_x {\partial {\cal L} \over \partial (\partial_x E)},
\label{derifun}
\end{equation}
with similar expressions holding for the other variables. From expression 
(\ref{derifun}) one obtains the complex conjugate of Eq. (\ref{cael}).
In terms of averaged Lagrangeans, what has to be done now is to substitute into 
Eq. (\ref{lagf}) the one-soliton solution, Eq. (\ref{shape}), plus the ion-acoustic 
field, Eq. (\ref{shapeion}). Doing this and performing the spatial integrations 
one arrives at
\begin{equation}
L \approx -2 \eta \dot \Phi + \left[ {4 W^2 \over 3 w} 
- {2 W \over 3 w^2} + 0.429 {W^2 {\dot w}^2 \over w} 
- 3.290 W \, {\dot w} \, w \, \dot A \right] + 
{\pi {\dot A}^2 \over k^3} - {\pi A^2 \over k},
\label{lagld}
\end{equation}
with $\eta = a(t)^2 \, w(t)$. The various numerical factors appear in 
Eq. (\ref{lagld}) as a result of the integrals involving trigonometric and 
hyperbolic functions. 

\begin{figure}[h]
\caption{Contour levels for the unperturbed dynamics $A \rightarrow 0$; 
$\eta=\sqrt{0.1/2}$.}
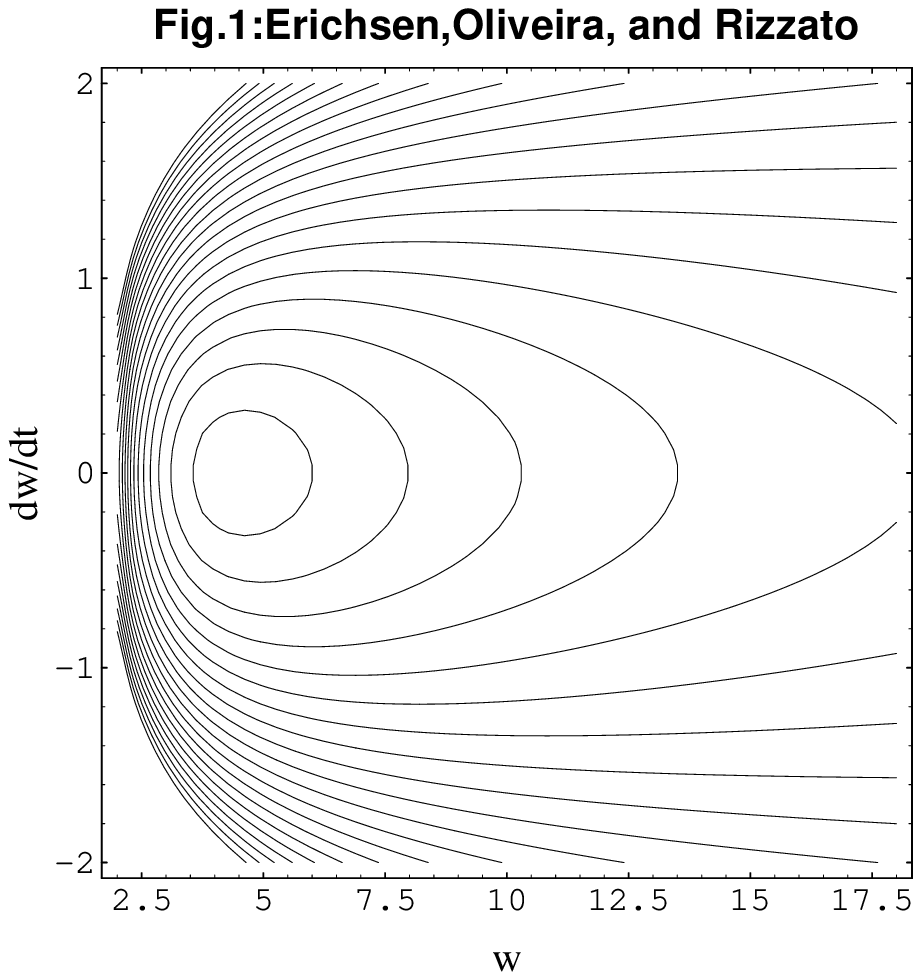
\label{fig1.ps}
\end{figure}

Euler-Lagrange equation with respect to the variable $\Phi$ indicates 
that $\eta$ is a constant of motion. As a matter of fact this feature has been 
already used to simplify the form of the Lagrangean (\ref{lagld}) by dropping 
terms proportional to $\dot \eta$ up to positive powers. 
Euler-Lagrange variational equations are then applied to the independent 
variables $w(t)$ and $A(t)$ to produce a two-degrees-of-freedom conservative 
dynamical system. If we set $A \rightarrow 0$ we have solutions corresponding 
to free oscillations of the soliton shape. One can construct a convenient 
phase-space to visualize those oscillations. This is done in Fig. (\ref{fig1.ps}) 
where we plot $\dot w(t)$ versus $w(t)$. The central fixed point of the figure is 
simply the static soliton solution analytically represented by 
Eq. (\ref{shapest}), and the curves surrounding the fixed point represent 
oscillatory modes of the soliton, each mode labeled by a particular 
constant energy that can be canonically evaluated from Lagrangean 
(\ref{lagld}) with $A=0$. In the absence of ion-acoustic free fluctuations, 
one can estimate the position of the fixed point, 
\begin{equation}
a_s = 1/w_s = \eta,
\label{largura}
\end{equation}
and the oscillatory frequency around the fixed point, 
\begin{equation}
\omega_s = \sqrt{2 \eta^2 \over 1.29} \sim a_s.
\label{frequsol}
\end{equation}
Given that $L_s \equiv w_s = 1/a_s$, one has $L_s \sim 1/\omega_s$, and given 
that $2 \pi/\omega_i = L_i \gg L_s$ one obtains a relationship 
involving the frequencies of soliton and ion-acoustic waves: 
\begin{equation}
\omega_i \ll \omega_s.
\label{relfreq}
\end{equation}
In other words, the components of the ion-acoustic field most weakly enslaved to 
the Langmuir field are those for which both length and time scales are 
much longer than the scales corresponding to the solitons.
One shall also mention that in addition to trapped orbits around the fixed 
point, open orbits are also possible. Those would represent decaying 
solitons for which $w \rightarrow \infty$ asymptotically. The fact that 
one has trapped and untrapped orbits implies that a separatrix 
does exist in which vicinity some amount of chaotic activity may be displayed 
if the system is in fact nonintegrable. The role of chaos, if chaos is indeed 
present, shall be better explored in the next section. 

\section{Full simulations versus the low dimensional model}

At this point we make use of the numerical techniques discussed in \S 2 
to compare results of full one dimensional simulations with the 
low dimensional model developed in the preceding section. 
Our full simulations give an account of the behavior of a stationary soliton 
submitted to the action of long wavelength ion-acoustic perturbations. Our 
purpose is to test the robustness of the soliton solution and see what 
happens when it looses stability due to the ion-acoustic radiation. Before 
embarking into the simulations it is perhaps convenient to preview the basic 
system behavior, based on possible results obtained with the low dimensional 
model. If the parameters of the low dimensional model are such that the 
corresponding nested orbits on the phase-plane $\dot w, w$ are mostly regular, 
one can expect a negligible influence exerted by the ion-acoustic field on the 
solitary structure in the full simulations, be this structure oscillatory or 
not. On the other hand it may well happen that the low dimensional system be 
nonintegrable. Should this be the case, and if low dimensional chaos is indeed 
well developed, the influence of ion-acoustic waves may be strong enough to 
destroy the solitary structure. In this case our low dimensional description 
may be expected to cease furnishing reasonable results. What is likely to 
happen then is that the chaotic low dimensional degrees-of-freedom start to 
act like a random drive, continuously delivering energy in a diffusive way to 
all the other dynamical modes \cite{lili91}. Then one may anticipate the 
soliton to decrease in intensity as its energy flows away. In addition, short 
wavelength modes are expected to grow and appear in the spectrum. We shall 
investigate some details of the transfer next.

\subsection{Low dimensional analysis}

\begin{figure}[h]
\caption{Poincar\'e plots ($\dot w, w$) of the low dimensional model 
with $k_i=0.0257$ and $\eta=\sqrt{0.1/2}$. $A_o = 0$ in (a), $0.14$ in (b), 
and $0.16$ in (c).}
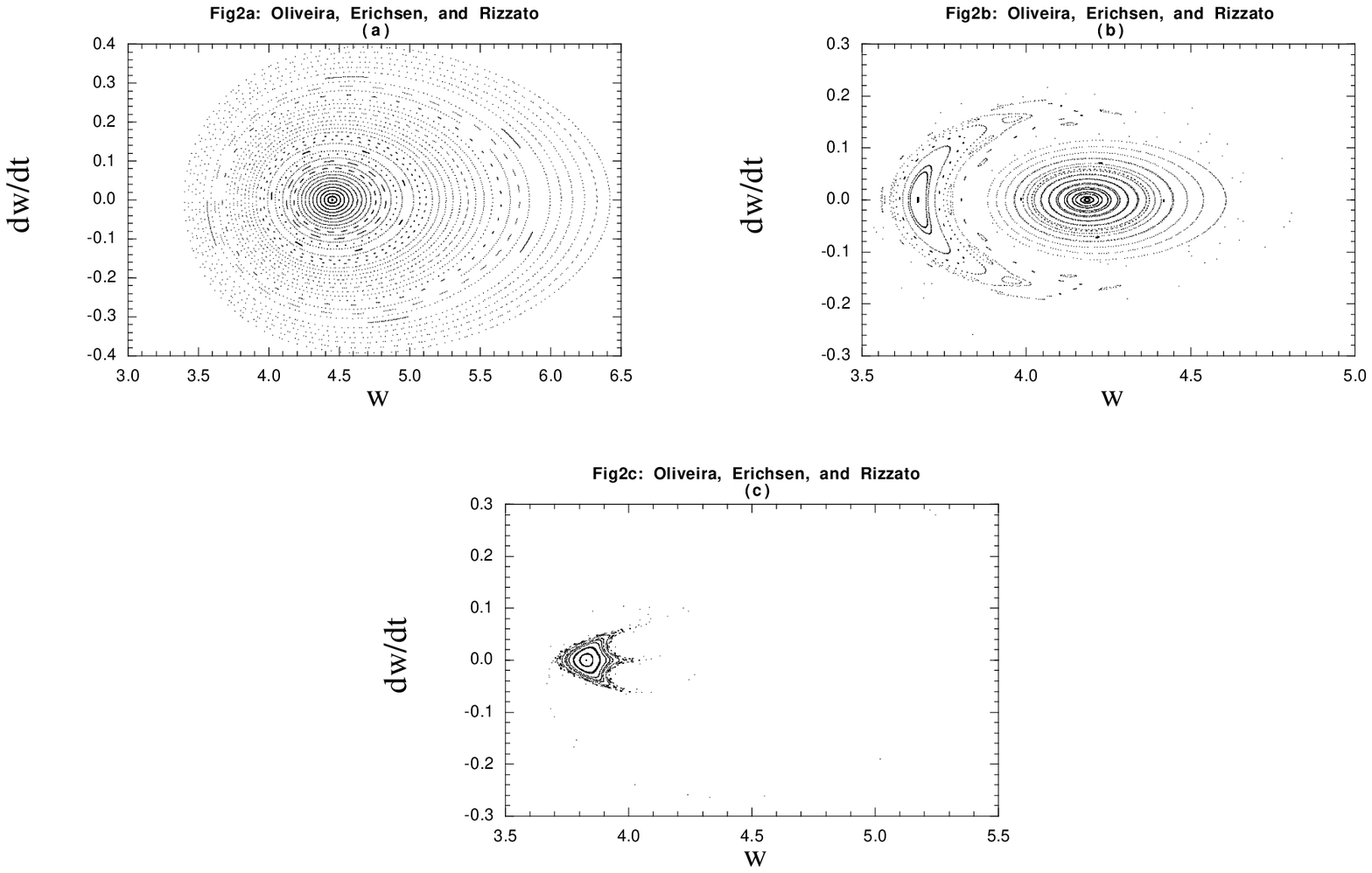
\label{fig2.ps}
\end{figure}

Let us first explore the integrability properties of the low dimensional 
approximation. To do that we examine the surface of section obtained when we 
record the pair of variables $w,\dot w$ each time $A = 0$ with $\dot A>0$. 
In the context of the low dimensional analysis we examine the system 
as an ion-acoustic 
wave of initial amplitude $A_o$ is added to the central fixed point of 
Fig. (\ref{fig1.ps}) - the remaining initial amplitudes corresponding to 
other orbits are obtained with help of the condition of constant energy. 
This constant energy is to be obtained from Lagrangean (\ref{lagld}). 
Parameters are specified in the legend of Fig. (\ref{fig2.ps}). In both low 
dimensional and full simulation we work with a soliton of $a_s = \sqrt{0.1/2}$ 
and with a perturbing ion-acoustic wavevector $k_i=0.0257$. For those 
parameters, 
\begin{equation}
L_i \sim 40 L_s,
\label{relcompl}
\end{equation}
and
\begin{equation}
\omega_s = 10.8 \omega_i.
\label{relfreql}
\end{equation}
One thus has $\omega_s \gg \omega_i$ and $L_s \ll L_i$ as required by the 
assumptions on time and length scales. Relation (\ref{relfreql}) 
in particular says that if the system is indeed nonintegrable, a period one 
island is likely to appear close to the central fixed point in the $w,\dot w$ 
phase-space. 

Examining the phase plots of Fig. (\ref{fig2.ps}) one sees that for small 
amplitudes the phase-space is mostly regular. However for larger values two 
features become noticeable: (i) the dynamics is indeed nonintegrable, and 
(ii) for large enough values of the amplitude, chaotic dynamics is dominant. 
In addition, for sufficiently large amplitudes (in the present case 
$A_o \sim 0.145$) the central fixed point undergoes an inverse tangent 
bifurcation and disappear along with the unstable fixed point of the 
period one island seen in Fig. (\ref{fig2.ps}b). All these features strongly 
suggest that the stochastic drive mechanism may be operative causing energy 
transfer into small spatial scales for moderately large values of the 
perturbation. This type of behavior is found for other choices of the 
ratios $\omega_i/\omega_s$ and $L_i/L_s$ as long as relations (\ref{relcomp}) 
and (\ref{relfreq}) are respected. 

\subsection{Full simulations}

The results of full simulations can be found in Fig. (\ref{fig3.ps}) where 
we make plots of the space-time history of the field $|E(x,t)|^2$, and of the average 
number of active modes versus time. The average number of modes is an auxiliary 
tool that can help to study details of energy transfer that are not 
particularly apparent in the space-time plots. The average number of modes is 
denoted by $\sqrt{<N_{L,i}^2>}$ for Langmuir and ion-acoustic fields, 
respectively, and defined according to the following \cite{thy81}:
\begin{equation}
< N_{L}^2 > \equiv {\sum_m m^2 |E_m|^2 \over \sum_m |E_m|^2}.
\label{averagelan}
\end{equation} 
\begin{equation}
< N_{i}^2 > \equiv {\sum_m' m^2 |n_m|^2) \over \sum_m'|n_m|^2}.
\label{averageion}
\end{equation} 
The primes in definition (\ref{averageion}) mean that the ion modes 
into which energy is initially placed are to be excluded from the summation. We 
do this simply to obtain clearer results. The problem is that since all the initial 
ion-acoustic energy goes into one single mode, the statistics becomes poor if 
we do not make the exclusion. No problems of that sort occur with the 
Langmuir field, as solitons already involve a statistically good number of modes.

We launch a solitary structure of shape given by Eq. (\ref{shape}), with a 
slight mismatch between $a(t=0)$ and $1/w(t=0)$ such that the soliton can 
oscillate initially: we choose $a(t=0) = \sqrt{0.1/2}$ and 
$w(t=0) = 1.2/a(t=0)$. For small enough values of the ion-acoustic 
perturbation Fig. (\ref{fig3.ps}a) shows that the solitary structure 
maintains its original amplitude without noticeable damping. It is seen from 
Fig. (\ref{fig3.ps}b) that for this perturbing amplitude the number of modes 
involved in the dynamics does not change significantly as time evolves. It 
should be noticed that the present modes are those used to construct the 
solitary structure. 

For larger values of the ion-acoustic amplitude, as in 
Fig. (\ref{fig3.ps}c), the soliton gradually damps away as time advances. 
Fig. (\ref{fig3.ps}d) shows that energy diffusion is now present, and 
that in the ion-acoustic field it is considerably much faster than in the 
Langmuir field. Diffusion in the Langmuir field becomes in fact almost 
imperceptible for even smaller perturbing amplitudes as we shall see a 
little later. Note that plateaus in the plots are formed when all the modes 
used in the simulation become involved in the dynamics - at this stage energy 
would be dissipated if we had added dissipation terms for large values of 
wavevectors. In the present case represented in Figs. (\ref{fig3.ps}c) 
and (\ref{fig3.ps}d), the central fixed point is still present in 
low dimensional phase-plots, as indicated by Fig. (\ref{fig2.ps}b). One can 
therefore think in terms of stochastic drive models to describe this type of 
regime \cite{lili91}. Although the orbits are chaotic, the presence of the 
central fixed point offers some resistance against rapid destruction of the 
low dimensional chaotic system. This low dimensional system might therefore 
last long enough to serve as a drive delivering energy to short wavelengths 
modes.

Now, if $A$ is large enough, the localized solitary structure is rapidly destroyed as 
indicated in Figs. (\ref{fig3.ps}e) and Figs. (\ref{fig3.ps}f). Energy is transferred to 
short wavelengths over short periods of time. We point out that this fast process 
occurs for ion fields intense enough to destroy the central fixed point of the 
low dimensional system, as indicated in Fig. (\ref{fig2.ps}c). In addition, 
diffusive time scales for both Langmuir and ion-acoustic fields become similar 
in this fast regime. Under such conditions the stochastic drive may not be a very 
appropriate concept since the life time of the solitary structure is too 
short.

We emphasize, therefore, that three distinct regimes appear to be present: 

(i) If the initial perturbation is small, typically $A \ll 0.1$, there is 
no diffusion whatsoever towards small length scales. 

(ii) For larger values of the perturbation, $A \sim 0.1$, diffusion is 
observed in both Langmuir and ion fields. But diffusion in the ion field is 
much faster. If one diminishes not too much the perturbing 
amplitude and reduces the observation time, diffusion in the Langmuir field 
becomes almost imperceptible although diffusion in the ion field can still 
be observed. This is what can be seen in Fig. (\ref{fig3.ps}g) where one 
considers a perturbing amplitude smaller, but of the same order of 
magnitude, than the one used in 
Fig. (\ref{fig3.ps}d). In general, within this range of perturbing amplitudes, 
the central fixed point of the low dimensional phase-space is still present. 
This could explain the persistence of the solitary structure seen in 
Fig. (\ref{fig3.ps}c). Since solitons are persistent and typically chaotic 
here, this regime is perhaps the most appropriately described by the 
stochastic drive. The oscillating low dimensional subsystem formed by the 
soliton and the ion-acoustic wave would excite the remaining modes of the 
system. 
As mentioned, diffusion is very asymmetric, being much faster in the 
ion-acoustic field. But on examining Eq. (\ref{caion}), it is not 
unreasonable to say that the Langmuir field term, appearing in the 
form $\partial_x^2 |E(x,t)|^2$ on the right-hand side, can act 
similarly to a source delivering energy to the ion field on the left-hand 
side. The source-like behavior would enhance diffusion in the ion-acoustic 
field. 

(iii) Finally, when the amplitude attains sufficiently large values, 
$A > 0.1$, fast diffusion takes place in both fields. In contrast to the 
preceding case, here the time scales for diffusion in both fields are similar. 
We point out that the central fixed point of the phase-plots no longer exists 
for this range of relatively large perturbing amplitudes. Again, this 
could explain the short life of the solitary structures, as seen in 
Fig. (\ref{fig3.ps}e). 

\section{Concluding remarks}

In this paper we examined the interaction of an ion-acoustic harmonic mode 
with a solitary wave of the Zakharov equations. Here the interest is 
to see how far can a solitary wave resist before it is destroyed by 
long wavelength radiation and how this destruction takes place. 
Although some recent works show that solitons can be stable structures 
even in nonintegrable environments \cite{deg74,kry80,dya89,isi94}, 
what we see here is that if chaos 
is strong enough in the low dimensional approximations, solitons are 
in fact destroyed and energy transfer towards small spatial scales 
takes place.

We have observed that the dynamics can be divided into three categories 
as a function of the amplitude of the initial ion-acoustic wave. Considering 
$\sqrt{2} \, a_s = \sqrt{0.1}$, for small amplitudes, $A \ll 0.1$, there is 
energy transfer neither in Langmuir nor ion-acoustic fields. For 
moderately large amplitudes, $A \sim 0.1$, diffusion is observed 
mostly in the ion-acoustic field, and for sufficiently large amplitudes, 
$A > 0.1$, diffusion is fast and equally present in both fields. In the 
intermediary regime one can think in terms of a stochastic drive delivering 
energy to modes with short wavelengths. The drive would be formed as a 
result of the chaotic, but persistent, low dimensional dynamics. 
Persistence follows because for not too large amplitudes 
the central fixed point of the low dimensional 
phase plots is still unaffected by the interaction, which means that solitons 
last long enough to serve as stochastic drives. While soliton turbulence may 
well describe the regime of intermediary amplitudes, it may not be quite 
appropriate to describe the regime of large perturbing amplitudes since solitons
readily damp away there. The characteristics of the stochastic drive are not 
easy to be obtained because the dynamics on the chaotic space 
$w(t),\dot w (t)$ is not pendulum-like. Therefore, some known results on 
pendulum-like settings \cite{lili91} cannot be directly used here. Details 
and comparisons with the full simulations are currently under study.

Recalling our initial question in this paper, the conclusion is that 
the interaction of isolated solitons and ion-acoustic radiation alone is 
capable of driving energy transfer.

   
\acknowledgments
This work was partially supported by 
Financiadora de Estudos e Projetos (FINEP) and Conselho Nacional de 
Desenvolvimento Cient\'{\i}fico e Tecnol\'ogico (CNPq), Brazil. Numerical 
computing was performed on the CRAY Y-MP2E at the Universidade Federal do 
Rio Grande do Sul Supercomputing Center.  
%

%
%
%
\begin{figure}[h]
\caption{$I\>(\equiv |E(x,t)|^2)$ and $\sqrt{<N^2>}$ from full simulations 
with $k_i=0.0257$ and $a(t=0) = \sqrt{0.1/2}$, $w(t=0)=1.2/a(t=0)$. $A=0.005$ 
in (a) and (b), $0.1$ in (c) and (d), $0.2$ in (e) and (f), and $0.05$ in (g).
Time has been normalized by a factor of $5000/100=50$ and space by a factor of 
$1024/32=32$.}
\label{fig3.ps}
\end{figure}
\end{document}